\documentclass[11pt,final,conference,letter]{IEEEtran}
\usepackage[utf8]{inputenc}
\usepackage[T1]{fontenc}
\usepackage[noadjust]{cite}
\usepackage{url}
\usepackage[dvipdfmx]{graphicx}
\usepackage{bmpsize}

\begin{document}

\IEEEoverridecommandlockouts

\title{A Critical View on CIS Controls}

\author{
	\IEEEauthorblockN{Stjepan Groš}
	\IEEEauthorblockA{
		Laboratory for Information Security and Privacy\\
		Faculty of Electrical Engineering and Computing, University of Zagreb\\
		Unska bb, 10000 Zagreb, Croatia \\
		E-Mail: stjepan.gros@fer.hr}
	\thanks{This paper is based on a lecture presented on Cyber Security Symposium co-organized by Croatian Ministry of Defense and Minnesota National Guard from August 18th - 25th, 2019 on island of Mali Lošinj, Croatia}}

\maketitle

\begin{abstract}
CIS Controls is a set of 20 controls and 171 sub-controls that were
created with an idea of having a list of something to implement so
that organizations can increase their security. While good in theory,
it is a big question of how viable this approach is in practice, and
does it really help. There is only a minor number of critical views 
of CIS Controls and since CIS Controls are marketed by two very
influential organizations they are very popular. Yet, there are
alternatives published by ISO, NIST and even PCI consortium. In this
paper we critically assess CIS Controls, assumptions on which they
are based as well as validity of approach and claims made in its
favor. The conclusion is that scientific community should be more
active regarding this topic, but also that more material is necessary.
This is something that CIS and SANS should support if they want to
make CIS Controls viable alternative to other approaches.
\end{abstract}


\section{Introduction}
\label{sec:intro}

\textit{CIS Controls} is a name for a set of 20 controls that are marketed
as having the best ratio between resources spent on security protection and gains
achieved by lowering the risk of being compromised. CIS Controls are
heavily marketed by Center for Internet Security which is the owner of
those controls, and SANS Institute that initiated the creation of 20
critical security controls and had ownership rights for some time under
the name of \textit{20 critical security controls}.

Due to marketing push by CIS and SANS, on the Internet there are a lot of
materials available related to these controls, including two courses
offered by SANS SEC566 \cite{sec566} and SEC440 \cite{sec440}. In general,
these materials largely explain what CIS Controls are, are copies of
information available from the documents produced by CIS \cite{cisweb}, or
deal with how to implement those controls in some environment.

In the same time, it is very hard to find critical review of CIS Controls
in general, and the scientific scrutiny regarding CIS Controls is missing.
This is worrisome for two reasons. The first one is the obvious ignorance
by scientific community of something used very intensively in practice,
and the second reason is the simple fact that no technology is a silver
bullet regarding security. To make a proper decision every decision maker
must have a balanced view on the technology, solution or whatever he or she
prepares to use.

The goal of this paper is to provide a critical view on CIS Controls and
try to place them within the whole ecosystem of different controls, and
security governance in general. We are not going to analyze specific
technical aspects of the controls, as this warrants a paper for itself,
but instead we'll analyze surrounding claims made about CIS Controls in
different documents, assumptions underlying CIS Controls, and the validity
of the approach taken by CIS Controls to secure IT systems.
Finally, we'll also give some recommendations regarding how to improve
identified shortcomings and also give some pointers of potential work that
can be done by scientific community.

The paper is structured in the following way:

\begin{itemize}
\item Section \ref{sec:overview} gives an overview of CIS controls, its history and related documents that form a security framework.
\item Section \ref{sec:criticalview} critically analyzes claims that can be found in the literature regarding CIS Controls.
\item Section \ref{sec:requiredwork} lists what should be done in order for CIS Controls to become more credible.
\end{itemize}

The paper finishes with the conclusions and future work outlook and
list of references.

\section{CIS Controls}
\label{sec:overview}

In this section we give a short overview of CIS Controls. We are not going into
depth since it is not necessarily for the understanding of the rest of
the paper. Only necessary basics are covered.

\subsection{Overview}

CIS Controls are defined and described in a document published by Center
for Internet Security \cite{ciscontrols}. At the time this paper was
written the latest version was 7.1. published in April, 2019. CIS
published several other guides in which they detail the application of
CIS Controls in some specific environments, like in cloud \cite{ciscontrolscloud},
Internet of Things \cite{ciscontrolsiot}, etc. But we will ignore those
as they are not important for the main point of this paper.

Each of 20 CIS Controls is further subdivided into sub-controls, and in total
there are 171 sub-control across all 20 controls. The 20 controls are divided
into three groups, i) basic, ii) foundational, and iii) organizational.
Since the number of sub-controls is quite large and it is hard to expect that
they are fit for everyone, starting from version 7 of CIS Controls
specification, \textit{implementation groups} were introduced, three groups in
total. In essence, the idea is that the smallest enterprises implement only
implementation group 1 while the largest enterprises implement all three
implementation groups.

The basic idea behind CIS Controls is that there is so much information
available on the Internet regarding protection of information systems that
it become contra-productive, basically making things worse, i.e. less
secure. This is what CIS in documents specifying controls calls \textit{fog
of more} \cite{ciscontrols}. So, the approach taken to remedy this situation
is to have a single source of information with cut-down version of proposed
controls to be implemented.

\subsection{History \& Development}

The only source of information about development of CIS Controls is a SANS
Institute Web page that contains narrative about its history \cite{cishistory}.
Searching on the Internet it is possible to find other sources too, like
e.g. \cite{ortmann2018} which is very high in Google search results, but
all of them basically repeat information from the SANS Institute Web page.
CIS Controls had a fore-funner but it was done for the US Government by the
US Government and thus it was not publicly released.

CIS Controls started
as an initiative by CIS and SANS when they contacted NSA with a proposal
to develop cut-down version of controls that will stop and detect most
attacks in existence. In due course a number of companies and institutions
joined the initiative started by SANS and CIS. Finally, in 2009. the first
version of CIS Controls was published. The current version is 7.1 published
in April 2019 as controls are reviewed on an annual basis.

The initial ownership on CIS Controls had SANS which called it \textit{20
critical security controls} but transferred it to Council on Cyber Security (CCS),
and finally in 2015 to Center for Information Security \cite{wiki:cis}.

In the historical overview of development of CIS Controls, SANS gives a
certain number of claims in favor of CIS Controls, but which are of
dubious value. We'll get to that later in the paper.

\subsection{Other Frameworks}
\label{sec:otherframeworks}

CIS Controls are not the only catalog of controls that can be found on the
Internet. There's ISO 27001 \cite{iso27001}, NIST Special Publication 800-53,
Revision 5 \cite{sp800-53r5}, NIST Cyber Security Framework v1.1 \cite{nistcf},
and PCI-DSS \cite{pcidss}.

ISO 27001 is part of the ISO 27000 series which is a comprehensive framework
for security management, thus something much larger than CIS Controls. In
Appendix A of ISO 27001 there are total of 114 controls grouped into 14 clauses
\cite{iso27001}.

NIST Special Publication 800-53r5 which is in draft status has 294 controls
grouped into 20 categories. This is twice more than the number of sub-controls
recommended by CIS.

PCI-DSS was created by card companies and, unlike others, is specific to financial
industry. It is created with a specific purpose to protect cardholder data. So, it's
narrower in scope than CIS Controls, and especially other frameworks.

The key difference between CIS Controls and other frameworks is that CIS Controls
are something you should implement, while all the others are something you 
choose from based on risk assessment you've done. In other words, in case of
CIS Controls someone already did risk assessment for you, have chosen controls
to implement and then handed over the list to you.

\subsection{Related Work}
\label{sec:relatedwork}

To the best of author's knowledge, the only critical view on the CIS Controls
is a blog post written by Ben Tomhave in 2011 \cite{tomhave2011}. The key claims
of his post regarding CIS Controls are:

\begin{enumerate}
\item they're not controls,
\item they're not scalable, and
\item they're designed to sell a product.
\end{enumerate}

The first claim revolves around the somewhat philosophical question of
what is \textit{control} and \textit{control statement}. The author believes
that for something to be control it has to be actionable, i.e. stated in
such a way to allow someone to do something specific or to perform some
action. Based on that premise the author argues that CIS Controls are
actually not controls but wish list. The second claim is about company
size, i.e. that those controls are not applicable to companies of all sizes
but only for the largest ones.  The basis for such thinking is, for example,
Data Loss Prevention suggested by CIS Controls is supposedly not applicable
for everyone. Finally, in the third claim Tomhave argues that the 20
critical controls advocate specific products, and even specific vendors.

In the same post Tomhave also draws attention to the fact that CIS Controls
miss policies, governance and risk management approach.


\section{Critical View on CIS Controls}
\label{sec:criticalview}

We can analyze CIS Controls from several different perspectives. The first
one is related to the whole idea of having certain number of selected
controls that everyone uses to improve security. This is done in subsection
\ref{sec:cisidea}. Related to the idea of having a specific number of
controls itself are assumptions, implicit and explicit, and context for
development of CIS Controls. This analysis is presented in subsection
\ref{sec:assumptions}. The third view concerns with question on how the
controls were selected, i.e. what was the process used to determine which
controls are better then the others. This is done in subsection
\ref{sec:development}. Finally, the fourth view is about the claims made
by CIS and SANS in favor of CIS Controls. Those are analyzed in subsection
\ref{sec:claims}.

Note that it is also possible to analyze controls themselves, as well as
recommendations given.  But this has been left for future work.

\subsection{Validity of CIS' Approach to Security}
\label{sec:cisidea}

By \textit{validity} of CIS' approach we mean three different things.

\subsubsection{Governance}

CIS Controls started as an idea of selecting a certain number of controls
that have the best ratio of resources invested into implementing controls
and security gains achieved, i.e. you would take the list and start to
implement everything on the list starting from the first one and finishing
with the last one.

The problem with this approach is how useful it is in practice with respect
to the governance of the security. To illustrate, let's say you were named
CISO in some company and your task it to make it secure. CIS Controls were
made with an assumption that you, as a CISO, don't ask anything but 
immediately start with implementing controls. But, as CISO, your first task
would be to assess current state of the security, which is done via risk
assessment process. This is in a contradiction with CIS Controls initial
idea as they assumed they did risk assessment for you.

Risk assessment is something very dependent on particular context, and there's
no way to have anything even remotely resembling universal risk assessment, i.e.
having something that will be reusable. It's always necessary to adjust to
specific environment. And, actually, the document about CIS Controls is specific
about that \cite{ciscontrols}:

\begin{quote}
But \textit{this is not a one-size-fits-all solution}, in either content or priority.
\textit{You must still understand what is critical to your business, data, systems, networks,
and infrastructures, and you must consider the adversary actions that could impact
your ability to be successful in the business or operations}. Even a relatively
small number of Controls cannot be executed all at once, so \textit{you will need to
develop a plan for assessment, implementation, and process management}.
\end{quote}

In other words, what this quote says is that you still need to do a risk management,
and in particular risk assessment process. Finally, in April 2018, CIS introduced
CIS RAM \cite{cisram} which means that the whole CIS Controls initiative was back on
what it tried to not be, i.e. risk management process.

With all this development in the CIS approached other frameworks, most notably
the ones specified by NIST.

\subsubsection{Target Users}

It is unclear to whom are these controls targeted, small, medium or large
enterprises. The fact is, those are vastly different, at least small and
medium enterprises on one hand and large enterprises on the other hand.
This was, in some sense, also recognized by CIS which published a separate
white paper detailing implementation of CIS Controls in SMEs \cite{cissme},
but also by introducing implementation groups in the version 7 of the
specification.

The bigger the organization is, the more planning and prioritization is
necessary. Planning and prioritization is a direct consequence of performing
risk assessment. Furthermore, to implement controls it is necessary to have
appropriate management structure which, with a number of users, increases.
CIS doesn't say anything about that, so this suggests that CIS Controls
aren't fit for large enterprises.

It could be argued that in large organizations there are already such
governance structures and that in such cases they could use CIS Controls
and manage the process. The problem with this argument is that, if there
are already appropriate governance structures, they probably already know
what to do, and besides, in such cases maybe ISO 27001 is more appropriate
to be used. Not to mention that in such cases there are probably no
problems with \textit{fog of more}.

\subsection{Assumptions underlying CIS Controls}
\label{sec:assumptions}

CIS Controls document is contradictory in that it lists security controls
that you can implement. And, in order to be able to do so, risk assessment
had to be performed by the people and organizations developing CIS Controls.
Namely, they assessed which vulnerabilities are the most critical ones, and
they then suggested which controls should be applied to lower the risk.

But then it warns you that,

\begin{quote}
\textit{... this is not a one-size-fits-all solution, in either content or priority.
You must still understand what is critical to your business, data, systems, networks,
and infrastructures, and you must consider the adversary actions that could impact
your ability to be successful in the business or operations. Even a relatively
small number of Controls cannot be executed all at once, so you will need to
develop a plan for assessment, implementation, and process management.}
\end{quote}

In other words, it allows you to skip over risk assessment process, and to jump
straight into risk control part of risk management process, and in particular
into reduction of the risk. But then, it suggests to you to develop a some kind
of a plan, which requires at least some steps from risk management process.

We have several problems there. \textit{The first} one is that CIS Controls
document in the same time gives you controls to implement and warns you that
you still need to assess your security. So, if implemented properly, how are
CIS Controls different that other risk management approaches that come with catalog of
controls? \textit{The second} problem is related more to how people behave,
that is, they don't read fine print. In other words, many people/organizations
will start to implement controls, without knowing is that good for their
specific situation or not.  CIS guide for SMEs \cite{cissme} goes even further
than that, it suggests using someone within the company that is otherwise not
related to security to guide the process of control implementation. This makes
thing even worse because security shouldn't be done by the people not knowing
security, and without proper governance structure - no matter how thin.

Also, quite explicit assumption is that you, as someone taking care of
security, are lost in lot of information, i.e. \textit{fog of more}. This
is actually debatable, since we might ask, if you, as a professional,
are lost in lots of information then are you professional at all, at least
security professional? Namely, one of the skills professionals have is to
know what is important and what is not, and more importantly, to build
their own methods and processes by which they work. So, \textit{fog of
more} doesn't sound like a real problem, at least not for security
professionals. Especially it doesn't sound as right approach to solve
it by giving people something simple instead of trying to teach them
how to handle that situation.

Anyway, turns out that CIS Controls could be beneficial to some small
number of companies that for whatever reason can not afford security
personnel (in-house or out-sourced) but need to deal with security
issues. At least that was the case until CIS published risk management
methodology \cite{cisram} which potentially made CIS Controls harder
for use for such companies.


\subsection{CIS Controls Development Process}
\label{sec:development}

As we already stated, the only source of information about the development
of CIS Controls available is a document describing how CIS controls were
developed. Unfortunately, that is far from enough for any objective
evaluation of CIS Controls. Furthermore, there is a number of claims and
arguments made in favor of CIS Critical Controls which are of dubious value.
We review them in the next subsection.

What is known about development process is the following \cite{cishistory}:

\begin{enumerate}
\item	We know a large number of organizations and people participated.
\item	Participants discussed a lot about which controls to include.
\item	There was a period of public consultations on which over 50 comments were received.
\end{enumerate}

The main argument for CIS Controls is given by emphasizing that a large
number of high-profile organizations participated in the development, like
UK's CESG and CPNI, the DoD chief computer network architect, etc. This
is a great example of \textit{appeal to authority} argumentation \cite{wiki:appeal}.
No serious technology should be based on who did it, but on how it was done.
In this case, more transparency in how things were organized, what decisions
were made, who did what, and so on would be much more useful that enumerating
participants.

When speaking about large number of people and organizations participating
in the development of something the question that arises is how they aggregated
data they received from participants.

To conclude, no written trace of development activities is available, at least not publicly.
This is problematic and should be corrected by publishing working papers
as well as reports that will allow scrutiny, and validation, of the process
and its outcomes.

\subsection{Claims made about CIS Controls}
\label{sec:claims}

When controls were selected and prioritized during development of CIS
controls, the guiding principle was that \textit{no control should be made
a priority unless it could be shown to stop or mitigate a \textbf{known attack}}
\cite{cishistory}. The question that immediately pops up is, what is
regarded as a \textit{known attack}? Attacks are different, and have their
own specifics. So, knowing attack at one organization doesn't mean we are
immediately knowing attacks in some other organization. What is probably
meant by the phrase known attack are \textit{techniques and tactics}
\cite{strom2017} that are used to perform attack which are much more
stable and common.


Next claim about CIS Controls is that it is the best practice \cite{ciscontrols}.
To analyze the validity
of this claim let's start with the definition of the term \textit{best practice}.
According to Merriam-Webster \textit{best practice} is defined as \cite{bestpractice}:

\begin{quote}
a procedure that has been shown by \textit{research} and \textit{experience}
to produce \textit{optimal results} and that is established or proposed as a
standard suitable for widespread adoption
\end{quote}

So, in order for something to be regarded as best practice, research and
experience has to show that it produces optimal results. There's no research
regarding CIS Controls, so we are left with experience. On the other hand,
the only experience available are anecdotal claims made in some of the
documents promoting CIS Controls. There are no documents recording
experiences and that makes this claim hard to fact check. So, it is
left as an open question if CIS Controls are best practice, or not.

Yet another claim about the usefulness of CIS Controls, which
additionally might support claim of being best practice, is the following
one \cite{cishistory}:

\begin{quote}
In June 2012, the Idaho National Laboratory, home of the National SCADA
Test Bed, of the U.S. Department of Energy, completed a very favorable
analysis of how the CIS Critical Controls applied in the electric sector as a
first step in assessing the applicability of the controls to specific industrial
sectors.
\end{quote}

It is very hard, almost impossible to find source of this claim. Additionally,
what is meant by \textit{favorable analysis}? How can that be checked, and
validated by doing the same analysis by someone else?

There are also another claim in favor of CIS Controls \cite{ciscontrols}:

\begin{quote}
In 2009, the U.S. Department of State validated the consensus controls by determining
whether the controls covered the 3,085 attacks it had experienced in FY 2009.
The State Department CISO reported remarkable alignment of the consensus
controls and the State Department actual attacks.
\end{quote}

This claim opens up more questions than it manages to give answers. How
this validation has been performed? What were results for each one attack?
What kind of attacks were those? How severe they were? What does \textit{alignment}
mean? How was alignment measured to be able to state that there is remarkable
alignment? No data has been published, and thus nothing can be checked about
these claims making them nothing more then a marketing speak.

Next, there is a statement about reduction of vulnerability based risk
which CIS Controls allow \cite{cishistory} which is widely copied around (e.g.
few top results from Google search \cite{cissuccess1}\cite{cissuccess2}\cite{cisreport}):

\begin{quote}
With a very rapid achievement of a more than 88\% reduction in
vulnerability-based risk across 85,000 systems, the State Department's
program became a model for large government and private sector organizations.
\end{quote}

This quote, and specific numbers mentioned, open up a number of questions.
First, what is \textit{vulnerability-based risk}? We might speculate that
it is risk calculated only based on known vulnerabilities, but it would be
better that this is clearly stated. Next, question is how did they calculate
or measure this reduction? Did they analyze all 85,000 systems, and if so, how?
To do anything across such a large number of systems is very demanding. Not
to mention that knowing what is there is also demanding.

To conclude, it is clear that a number of claims is made in favor of CIS
Controls, yet none of these claims can be independently verified. Also
very important is that security is a fast moving field, what was valid
few years ago, doesn't have to be valid today. So, even though CIS has
annual reviews of CIS Controls, all the claims about it are not revalidated.

\section{What needs to be done}
\label{sec:requiredwork}

So, it is obvious that further work is necessary that will evaluate true
usefulness of CIS Controls. But, we should be clear from whom we might expect
objective evaluation. For example, there are case studies available on the
CIS Web page \cite{ciscasestudies}, but all of them are primarily marketing
type of materials meant to persuade someone to use CIS Controls. There is
one report written as a part of GIAC (GCCC) Gold Certification in 2016
\cite{bosco2016}, but it is also not enough.

We argue that more scientific analysis is necessary in order to make CIS
Controls more viable alternative for making systems more secure. To get
to that conclusion it is necessary to analyse incentives of stakeholders
in the ecosystem built around CIS Controls. The stakeholders in this system
are:

\begin{itemize}
	\item Center for Internet Security (CIS) -- who has ownership of CIS Controls,
	\item SANS -- who started work on CIS Controls and also gives courses that teach CIS Controls,
	\item solution vendors -- that offer solutions for the implementation of CIS Controls, and
	\item users -- who implement CIS Controls in order to protect themselves.
\end{itemize}


For each stakeholder, the following questions should be asked:

\begin{itemize}
	\item What they can gain/lose by criticizing CIS Controls?
	\item What they can gain/lose by ignoring CIS Controls?
	\item Do they have incentive to thoroughly analyze CIS Control benefits?
\end{itemize}

No stakeholder has an incentive to contradict validity and usefulness of CIS
controls, or to say they are no better then other approaches. CIS and SANS
started everything and to suddenly claim it doesn't work would be disastrous
for them. In addition, SANS has very expensive courses built on CIS Controls.
Solution vendors will happily sell anything customers believe works, and have
absolutely no incentive to say it doesn't help, or that it helps a lot less
than people think. Users might be an exemption, but people tend not to
criticize something they did (otherwise, they did it for no purpose).
Also, users tend to rely on vendors' opinion and what they can find on the Internet.
Finally, users tend to believe authorities and seek protection from authorities.
Namely, if incident happens even though CIS Controls are implemented, it will
have repercussions less likely since \textit{everybody} is doing it.

The conclusion is that CIS Controls have to be validated by someone not
having a stake in them, and that ones are scientist that should perform
scientific analysis. But scientist can not analyze something that is not
described, so scientist should be given access to specific CIS Controls
implementations, or at least there should be better case studies published.

So, the following work is necessary to perform in order to make CIS Controls more
credible:

\begin{itemize}
	\item Papers describing experience/case-studies with the implementation of CIS Controls
	\item Describe context in which CIS Controls were decided to be used
	\item What controls and in which order were they implemented?
	\item How well were they implemented? What problems were encountered?
	\item How effective they are? Were there breaches?
	\item Surveys and reviews of experience papers
	\item Synthesis and generalization of results presented in the case-studies
	\item Critical assessment of experience/case-studies papers
\end{itemize}

\section{Conclusions and Future Work}
\label{sec:conclusions}

CIS Controls started as a quest for the simple solution that will solve majority
of security problems for everyone. The solution should be preferably plug-and-play,
or at most “Next-Next-Finish”. And for some time it was marketed as such even
though there was fine print about knowing your system before doing anything.
But as time progressed, and the whole idea evolved about having ready-made controls to
implement evolved it ended up being one more risk management solution with
accompanying catalog of controls to be chosen from, only this time (maybe) somehow reduced.


The fact is that cyber security and information security are not easy, no matter
how much we tried to make them such. In part this is because they are
very context dependent, which make it questionable if a one-size-fits-all solution
will ever exist.

CIS Controls are heavily marketed by the simple fact that CIS and SANS are very
influential organizations, and in this marketing certain statements are made that
are vague, can not be checked, or are simply wrong.

In the end, any protection is better than none, but also not all protections are
equally good. So, deeper analysis is necessary of gains achieved by using CIS
Controls when compared to other options, and in the end when no protections are
used at all in some structured way. We suggested some tasks that could be done
in order to improve things, but much is on CIS and SANS.

\IEEEtriggeratref{12}

\bibliographystyle{IEEEtran}
\bibliography{bibliography}

\end{document}